\documentclass[fleqn,usenatbib,useAMS]{mnras}

\usepackage{amsmath,amssymb}
\usepackage{threeparttable}
\usepackage{hyperref}
\usepackage{fontawesome5} 
\usepackage{lineno}

\newcommand{\orcid}[1]{\textsuperscript{\href{https://orcid.org/#1}{\faOrcid}}}

\usepackage{graphicx}	
\usepackage{amsmath}	
\usepackage{multicol}        
\usepackage{bm}		
\usepackage{pdflscape}	

\usepackage[T1]{fontenc}
\usepackage{ae,aecompl}

\usepackage{newtxtext,newtxmath}

\title[Coronal fine dynamics in recurrent jets]{The fine dynamics in homologous and recurrent jets induced by persistent rising loops and mini-filaments}

\author[H. Wei et al.]{Hengyuan Wei\orcid{0000-0002-0210-6365}$^{1,2}$,
Zhenghua Huang\orcid{0000-0002-2358-5377}$^{1,2}$
\thanks{huangzh@nju.edu.cn},
Yadan Duan\orcid{0000-0001-9491-699X}$^{3,4}$,
Chuan Li\orcid{0000-0001-7693-4908}$^{1,5,6}$
\\
$^{1}$Institute of Science and Technology for Deep Space Exploration, Suzhou Campus, Nanjing University, Suzhou 215163, Jiangsu, China\\
$^{2}$State key Laboratory of Lunar and Planetary Sciences, Macau University of Science and Technology, Macau, China\\
$^{3}$Yunnan Observatories, Chinese Academy of Sciences, Kunming, 650216, China\\
$^{4}$Yunnan Key Laboratory of Solar Physics and Space Science, Kunming, 650216, China\\
$^{5}$School of Astronomy and Space Science, Nanjing University, Nanjing 210023, China\\
$^{6}$Key Laboratory of Modern Astronomy and Astrophysics (Nanjing University), Ministry of Education, Nanjing 210023, China
}

\date{Last updated 2020 June 10; in original form 2013 September 5}

\pubyear{2020}

\begin{document}
\label{firstpage}
\pagerange{\pageref{firstpage}--\pageref{lastpage}}
\maketitle

\begin{abstract}
Jets are common eruptive phenomena in the solar atmosphere which may occur repeatedly.
Many studies of their fine dynamics have been conducted.
 However, the fine dynamics of persistent interactions among various features that drive recurrent coronal jets have not been studied in detail.
In this paper, we use observations from the Solar Orbiter to report persistent interactions between rising loops/mini-filaments and a fan-spine-like structure, which produced more than 22 ejections.
Many loops and mini-filaments under the fan-spine-like structure rose with speeds of $8 - 58\,\mathrm{km~s^{-1}}$ and an average of $27\,\mathrm{km~s^{-1}}$.
These rising loops and mini-filaments interacted with the fan-spine-like structure successively, producing ejections with speeds ranging from 25 to 186\,\rm km\,s$^{-1}$ and an average at $80\,\rm km~s^{-1}$.
We observed the fine dynamics of the drivers of these recurrent jets in detail, including partial eruption of mini-filaments, formation of a new mini-filament by contraction of remaining threads from the partially-erupted mini-filament, and interaction between rising loops (or mini-filaments) and the fan-spine-like structure.
Brightenings appeared near the footpoint of these rising structures, followed by the formation of current sheets.
Some arcades at the outflow region contracted with speeds of around $10\,\rm km~s^{-1}$, and the outflow region moved at around $8\,\rm km~s^{-1}$ toward the opposite direction.
Bright blobs were observed in the current sheets, and they propagated at speeds averaging at $21\,\rm km\,s^{-1}$ and had an average width of 296\,km.
We emphasize the vital roles of persistent rising loops and/or mini-filaments in producing recurrent jets by interacting with the fan-like structure, and show their detailed dynamics with unprecedentedly-high-resolution observations.
\end{abstract}





\section{Introduction}

Jets are highly dynamic and eruptive phenomena in the solar atmosphere.
Hot coronal jets have been widely studied since the Skylab and Yohkoh eras with observations in X-rays and EUV passbands.
They consist of plasma ejected along open magnetic fields or large-scale loops and they are often observed in multiple passbands, such as X-ray, ultraviolet, H$\rm\alpha$, and He\,{\footnotesize I}\,\citep[e.g.][etc.]{1992PASJ...44L.173S,1995Natur.375...42Y,2012A&A...548A..62H,2013ApJ...769..134M,2019Sci...366..890S,2023A&A...672A.173W}, which indicates their multi-thermal nature.
They usually have a shape of inverted-Y or cusp in observations \citep[e.g.][etc.]{2011ApJ...735L..18L,2014Sci...346A.315T,2018ApJ...854...92T,2019ApJ...883..115N}.

Many previous studies have performed detailed analyses of the fine dynamics in triggering a single jet.
\citet{2012RAA....12..573C} identified multiple features propagating along a jet.
\citet{2014A&A...562A..98C} observed plasma flows propagating along the loops at the base of a jet before its birth.
\citet{2015Natur.523..437S} demonstrated that many jets formed due to eruptions of small-scaled filaments.
\citet{2018ApJ...854..155K} found the evidence of multiple phases of the erupted filament in a coronal hole jet and the activity sequence followed the breakout model.
\citet{2018ApJ...854...80H} observed multiple brightenings in the loops at the bottom of a jet and suggested that they were the results of magnetic braiding.
\citet{2022A&A...664A..28M} found that multiple plasma blobs propagated toward the dome apex, followed by high speed outflows at the jet spire.
\citet{2023ApJ...944...19L} provided a very detailed analysis of the trigger of a jet, which was related to the breakout reconnection between an erupting mini-filament and overlying loops followed by further reconnection along the erupting mini-filament.
\citet{2023A&A...673A..83L} found that bright ribbons formed under an erupting mini-filament and dark materials were ejected when the erupting mini-filament interacted with overlying loops, producing multiple bright blobs at the interaction region.
\citet{2024A&A...690A..11K} revealed the possible areas of energy release that was responsible for triggering the jet.

In some cases, jets occur repeatedly and originate from the same region, and this phenomenon is called ``recurrent jet'' or ``homologous jet''.
Many studies about the dynamics of recurrent jets have been conducted in the past decades.
\citet{2008A&A...491..279C} demonstrated that the recurring  X-ray jets were related to X-ray bright points and Ca II H brightenings, and the pre-jet loops expanded and erupted in the jet occurrence.
\citet{2008A&A...481L..57C} found a density dependence on velocity in high-velocity up-flows, which favours the evaporation scenario of jets.
\citet{2013A&A...555A..19G} demonstrated that the recurrent jets were related to the periodic build-up of electric currents, which were corresponding to elongated patterns at the photosphere.
\citet{2015ApJ...815...71C} performed detailed analysis of recurrent jets and found they were induced by continuous reconnection between a satellite spot/moving magnetic features and pre-existing fields.
\citet{2014A&A...567A..11Z} found that bright blobs formed during the occurrence of recurring EUV jets.
\citet{2025A&A...702A.201L} confirmed continuous magnetic cancellation at the footpoints of recurrent jets, and they observed the complex evolution of the fan-spine structure, which resulted in broad and fragmented outflows.
\citet{2025ApJ...979...62Z} studied the relationship between several physical quantities (total unsigned vertical current etc.) and the periodicity of jets, and demonstrated that the energy and helicity accumulated during the emergence, shearing and cancellation of magnetic fields are important in triggering recurrent jets.

The previous studies performed detailed analyses of the trigger mechanism of recurrent jets and dynamics during these jets. 
However, direct observations of the fine dynamics in interactions between rising structures and fan-spine structures are rare due to insufficient resolution, especially in the coronal observations.
In this paper, we use the unprecedentedly-high-resolution observations to make detailed analysis about the fine dynamics in persistent interactions between the rising loops/mini-filaments and a fan-spine-like system, which induced a series of recurrent jets.
We analysed the fine dynamics of how a mini-filament interacted with the fan-spine-like structure, partially erupted and induced a jet in detail, while its remaining threads contracted and formed a new mini-filament.
We also showed the fine dynamics in the interaction process between rising loops and the fan-spine-like structure, the fine process in the formation of current sheet and the fine dynamics in the outflow region.
In the following, we will introduce the observations from the Solar Orbiter, present the detailed studies about these recurrent jets and show the analysis of two jets for example in Section \ref{sec:res}.
The discussion and conclusion will be given in Section \ref{sec:rescon}.

\section{Observations and Results} \label{sec:res}

\begin{figure*}
\includegraphics[trim=0cm 0.2cm 0cm 0.5cm, width=\textwidth]{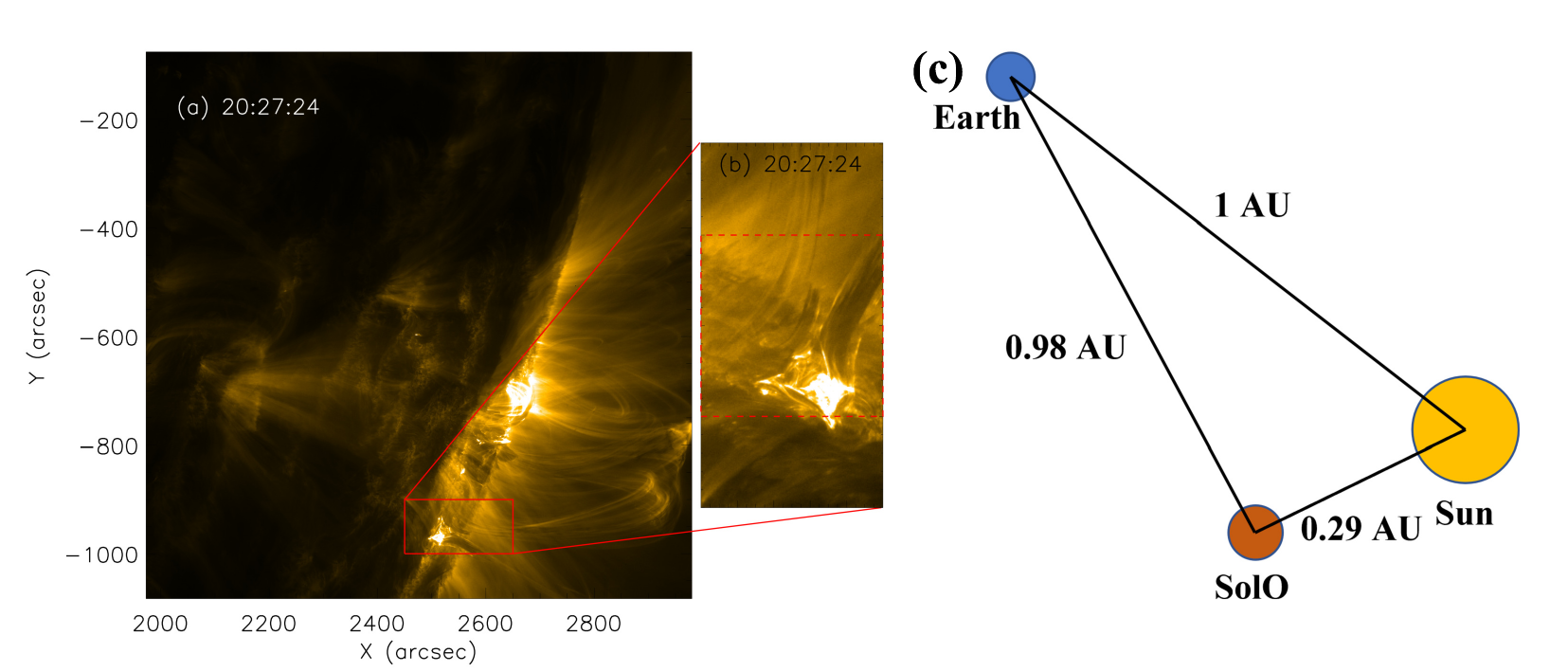}
\caption{Overview of the region of interest. 
(a) The whole field of view of EUI at 20:27:24\,UT.
The red rectangle encircles the region we studied.
(b) The zoomed-in snapshot of the region in red rectangle at the same time.
The snapshot in (b) is rotated 90 degrees counter-clockwise.
The red dashed rectangle in (b) marks the region used to produce the animation.
(c) The relative position of the Sun, Earth and SolO.
The evolution of the studied region (marked by red dashed rectangle in (b)) from 19:59:56\,UT to 23:55:08\,UT can be found in the associated animation.
The black arrows in the animation point to the structures associated with trigger processes of the jets and the white arrows in the animation mark the jets.
\label{fig:ove}}
\end{figure*}

The data analysed in this paper were observed by the Extreme Ultraviolet Imager \citep[EUI,][]{2020A&A...642A...8R} aboard the Solar Orbiter \citep[SolO,][]{2020A&A...642A...1M}.
The observations were taken from 19:59:56\,UT to 23:35:08\,UT on 2024 April 5 and retrieved from the EUI data release 6.0.
All the data are publicly available and given as reduced level 2 data.
The observations were taken by the High Resolution Imager (HRI), focusing on the southwestern hemisphere, and the event studied here occurred near the southwestern limb (encircled by a red rectangle in Figure \ref{fig:ove} (a)).
The SolO was 0.29 AU away from the Sun when it was observing, with a distance of 0.98 AU to the Earth (see Figure \ref{fig:ove} (c)).
The angular resolution of the HRI data is around 0.492\,\arcsec\,per pixel, which corresponds to around $100\,\rm km$ per pixel.
The temporal resolution of this data is $16\,\rm s$.

In this event, we observed a series of recurrent jets originating from the same fan-spine-like system on 2024 April 5 from 19:59:56\,UT to 23:35:08\,UT.
The fan-spine-like system can be seen in Figure \ref{fig:ove} (b), with many loops at the bottom and open fields (indicated by jets) at the surroundings.
These jets showed a canonical jet morphology, characterized by brightening at the jet base followed by the ejection of cool and/or hot plasma, which has been reported in many previous studies (see \citet{2011ApJ...731L...3S,2015Natur.523..437S} for example and more summarised in reviews by \citet{2016SSRv..201....1R}, \citet{2021RSPSA.47700217S}, \citet{2022FrASS...920183S} and \citet{2022AdSpR..70.1580S}).
We found more than 22 recurrent jets in this region and compiled statistics of their physical parameters and occurrence positions.
The tally of their parameters and occurrence positions are shown in Table \ref{tab:para}.
All these jets and their related trigger structures are marked by arrows and can be found evidently in the associated animation of Figure \ref{fig:ove}.
For many jets in our observations, they showed complex topology and features, such as the multiple brightenings at the top of the fan-like structure.

We observed that there were always some structures (including arcades and mini-filaments) rising from the bottom of the fan-spine-like system before the jets occurred, except several jets whose evolution at the base can not be clearly observed.
We can trace the rising trajectories of these structures and their fine dynamics during these jets easily in the relevant animation of Figure \ref{fig:ove}.
In table \ref{tab:para}, we found that 11 in 22 jets were triggered by rising loops, and these jets were dominated by bright ejections.
Two in 22 jets were facilitated by rising mini-filaments, and they were composed of dark ejections.
The remaining jets were unable to identify their trigger structures.
It should be noticed that these jets may be triggered by any of various dynamic fine structures, including mini-filaments, rising chromospheric fibrils, expanding magnetic domes, emerging toroidal flux tube or small-scale flux ropes as reported previously in both observations and numerical experiments\,\citep[e.g.][etc.]{2013ApJ...769L..21A,2013ApJ...771...20M,2015A&A...573A.130P,2015ApJ...798L..10L,2017Natur.544..452W,2018ApJ...852...98W,2018ApJ...864..165W,2021ApJ...921L..20P,2022ApJ...935L..21N,2023ApJ...943...24P,2023ApJ...958L..38N,2025A&A...699A..87P,2025ApJ...986...47Z}, because the present high-resolution data have only one passband and they may be not visible in EUI 174\,\AA.
The velocities of the rising structures were estimated by either the time-distance plots along their rising trajectories or the distance they moved divided by the time they spent.
The rising velocities of these structures fell in the range of $8 - 58\,\rm km~s^{-1}$, and there was no particular velocity range in which the rising speeds were in favours.
The rising loops/mini-filaments interacted with the fan-spine-like system in their following evolution, forming features like bright thin sheets under the jets and brightenings at the interacting sites, and the recurrent jets were produced after these interactions.

We estimated the velocities of these jets through either time-distance plots along their trajectories or their lengths divided by the time they passed through.
From table \ref{tab:para}, we found that the velocities of these jets were at the range of $25 - 186\,\rm km~s^{-1}$, with most velocities falling in the range of $40 - 110\,\rm km~s^{-1}$.
We also measured the lengths and widths of these jets.
The lengths of these jets ranged from $3\,\rm Mm$ to $18.6\,\rm Mm$ concentrating at around $4 - 9\,\rm Mm$, and the widths of these jets varied from $0.3\,\rm Mm$ to $3.6\,\rm Mm$ with $0.3 - 0.6\,\rm Mm$ being the favourite.
It is also noticeable that most jets occurred at the spire of the fan-like structure and only 4 in 22 jets happened at the lateral boundary of the fan-spine-like structure.
We also roughly estimated the kinetic energy of these jets through the equation $E = \frac{1}{2}mv^2$, assuming the density of the jets was around $10^{9}\,\rm cm^{-3}$\,\citep{2012A&A...545A..67M}, and we found that the kinetic energy was around $10^{21} - 10^{25}\,\rm erg$.

\begin{figure*}
\includegraphics[width=\textwidth]{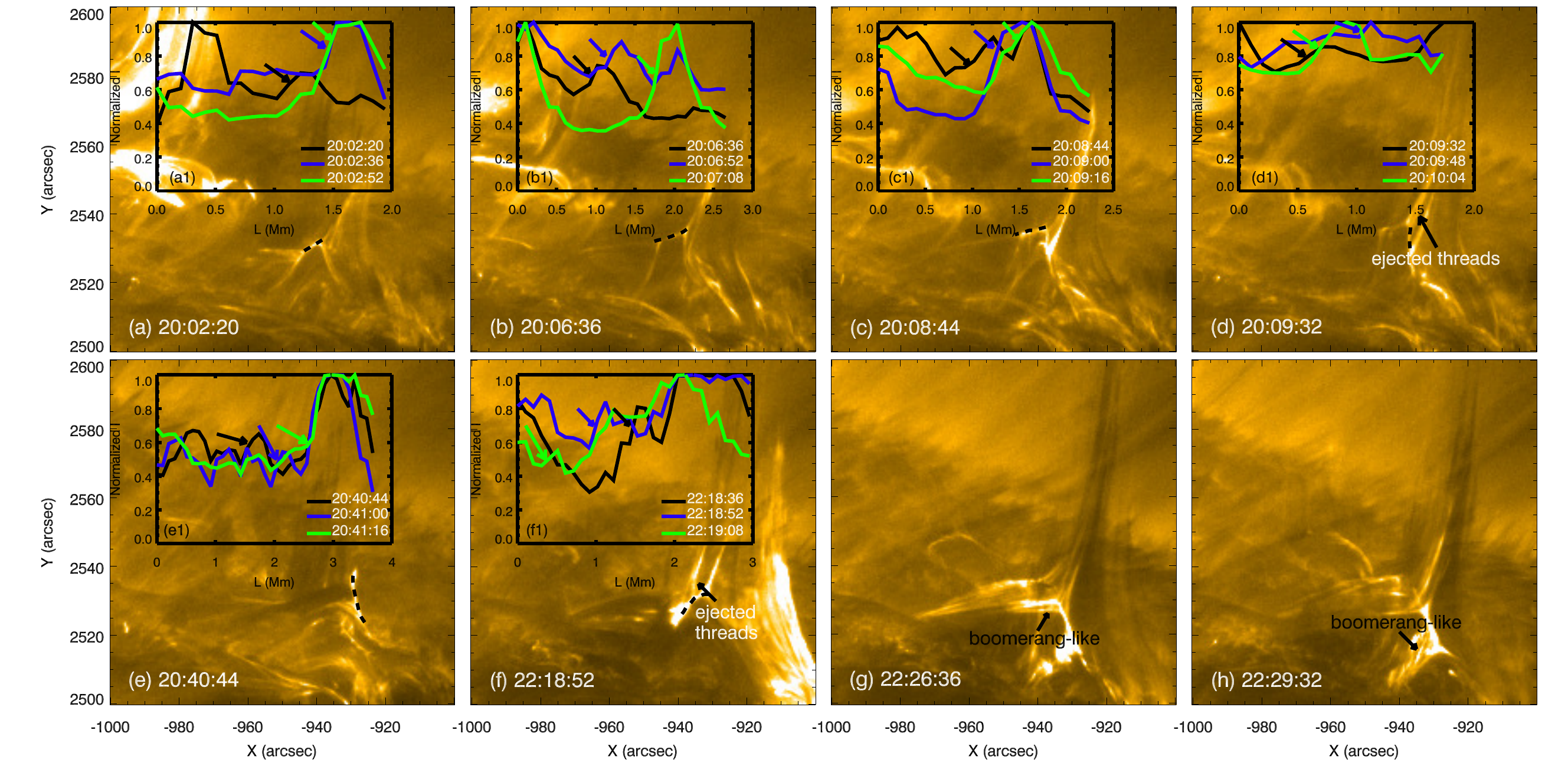}
\caption{The snapshots of the region of interest taken by EUI at 20:02:20\,UT, 20:06:36\,UT, 20:08:44\,UT, 20:09:32\,UT, 20:40:44\,UT, 22:18:52\,UT, 22:26:36\,UT and 22:29:32\,UT.
The dashed lines in (a) - (f) show the slits along which we obtain the intensity distributions.
The black arrows in (d) \& (f) point to the ejected threads originating from the bright blobs and parallel to the jets.
The black arrows in (g) \& (h) point to the boomerang-like structures at the outflow region of the reconnection.
Panels (a1) - (f1) show the intensity distribution along the slits in (a) - (f) respectively.
All the snapshots are rotated 90 degrees counter-clockwise as same as Figure\,\ref{fig:ove} (b).
\label{fig:pla}}
\end{figure*}

We found that bright blobs formed in and propagated along bright sheets under these jets, which can be clearly seen in the intensity distribution along the sheets at different time as shown in Figure \ref{fig:pla} (a1) - (f1) (pointed by arrows with different colours indicating different time).
The bright blobs in panels (a), (b), (c), (e) and (f) of Figure\,\ref{fig:pla} propagated unidirectionally along the bright thin sheet with speeds of $23\,\rm km~s^{-1}$, $15\,\rm km~s^{-1}$, $8\,\rm km~s^{-1}$, $28\,\rm km~s^{-1}$ and $28\,\rm km~s^{-1}$, respectively, and their widths were at $282\,\rm km$, $328\,\rm km$, $196\,\rm km$, $398\,\rm km$ and $161\,\rm km$.
The bright blob in panel (d) stayed in a similar place during its evolution (indicated in Figure \ref{fig:pla} (d1)), and its width was at around $410\,\rm km$.
The size of these blobs is consistent with those reported in \citet{2025A&A...702A.188N} while their propagating speeds are slightly lower.
We notice that these bright blobs showed different dynamics during their propagation.
The bright blobs shown in panels (a), (b), (c) and (e) only moved along the bright thin sheet without ejecting any plasma parallel to the jets, while those shown in panels (d) and (f) ejected plasma parallel to the jets and became a part of the jets later (see the black arrows in Figure \ref{fig:pla} (d) \& (f)).
We also observed some boomerang-like structures in the outflow regions of reconnection even when the individual blobs were not clearly visible (pointed by black arrows in Figure \ref{fig:pla} (g) \& (h)), and this may indicate plasmoid-mediated reconnection as shown by \citet{2025A&A...702A.188N}.

Two in 22 jets occurring at 20:21:48\,UT and 22:24:44\,UT were more violent and we are easier to trace the fine dynamics during their trigger processes, thus, we perform detailed analyses for them in the following subsections and take them as examples.

\begin{table*} 
    \centering 
    \footnotesize 
    \caption{Parameters of jets and rising structures}
    \label{tab:para}
    \begin{threeparttable}
        \begin{tabular*}{\textwidth}{@{}l l l r r r l l r l @{\extracolsep{\fill}}}
            \hline\hline 
            Jet\tnote{a} & Start Time & Location\tnote{b} & Length & Width & Speed\tnote{c} & Rising Structure\tnote{d} & Start Rising Time & Rising Speed\tnote{c} & Interaction position\tnote{e} \\
            & (UT) & ($X\arcsec,Y\arcsec$) & (km) & (km) & ($\rm km~s^{-1}$) & & (UT) & ($\rm km~s^{-1}$) & \\
            \hline 
            Jet1  & 20:01:00 & -936.9,2530.5 & 5825  & 427   & 65  & Loop     & 19:59:56 & --- & Spire      \\
            Jet2  & 20:06:04 & -926.8,2515.2 & 9093  & 307   & 74  & Loop     & 20:05:48 & --- & Boundary   \\
            Jet3  & 20:06:36 & -917.7,2513.8 & 12237 & 500   & 87  & ---      & ---      & --- & Boundary   \\
            Jet4  & 20:07:40 & -937.3,2528.0 & 4250  & 502   & 59  & Loop     & 20:06:52 & 39  & Spire      \\
            Jet5  & 20:21:48 & -930.7,2501.3 & 16669 & 3574  & 186 & Filament & 20:20:28 & 58  & Spire      \\
            Jet6  & 20:26:04 & -931.6,2518.0 & 18602 & 451   & 101 & ---      & ---      & --- & Boundary   \\
            Jet7  & 20:41:48 & -929.1,2535.4 & 8187  & 500   & 85  & Loop     & 20:41:00 & 48  & Spire      \\
            Jet8  & 20:43:40 & -928.5,2540.0 & 7053  & 502   & 86  & ---      & ---      & --- & Spire      \\
            Jet9  & 20:54:04 & -937.0,2529.1 & 13407 & 1410  & 26  & Filament & ---      & --- & Spire      \\
            Jet10 & 21:08:44 & -932.1,2525.5 & 5261  & 373   & 55  & ---      & ---      & --- & Spire      \\
            Jet11 & 21:24:28 & -933.1,2527.2 & 10712 & 1624  & 27  & ---      & ---      & --- & Spire      \\
            Jet12 & 21:41:00 & -930.4,2525.4 & 6687  & 500   & 139 & Loop     & 21:40:44 & 14  & Spire      \\
            Jet13 & 21:48:28 & -939.2,2529.8 & 6957  & 607   & 123 & Loop     & 21:47:40 & 10  & Spire      \\
            Jet14 & 21:51:56 & -932.9,2527.7 & 7708  & 478   & 107 & ---      & ---      & --- & Spire      \\
            Jet15 & 22:18:04 & -910.9,2505.9 & 15399 & 3462  & 107 & Loop     & 22:17:16 & 31  & Boundary   \\
            Jet16 & 22:18:20 & -940.0,2524.5 & 6390  & 485   & 90  & ---      & ---      & --- & Spire      \\
            Jet17 & 22:18:20 & -931.1,2532.5 & 5283  & 590   & 56  & ---      & ---      & --- & Spire      \\
            Jet18 & 22:24:44 & -933.8,2529.8 & 13264 & 590   & 81  & Loop     & 22:22:20 & 35  & Spire      \\
            Jet19 & 22:36:28 & -927.6,2527.7 & 11246 & 451   & 33  & Loop     & 22:35:08 & 10  & Spire      \\
            Jet20 & 22:45:32 & -922.3,2534.5 & 4015  & 384   & 42  & Loop     & 22:42:36 & 19  & Spire      \\
            Jet21 & 22:51:40 & -936.2,2525.2 & 3081  & 828   & 25  & Loop     & 22:49:16 & 8   & Spire      \\
            Jet22 & 23:22:20 & -933.1,2524.1 & 2954  & 308   & 92  & ---      & ---      & --- & Spire      \\
            \hline\hline 
        \end{tabular*}
        \begin{tablenotes}[flushleft, para]
            \footnotesize
            \item[a] Jets mentioned here refer to different ejections, and it is hard to identify whether they are all distinct jets due to lack of magnetic field information.\\
            \item[b] The locations of the bases of these jets.\\
            \item[c] The speeds of these jets and rising structures are obtained from either time-distance plots or their lengths divided by the time they passed through.\\
            \item[d] The rising structures marked by loops only demonstrate that we do not observe obvious mini-filaments, and it does not mean there must not exist mini-filaments, because we only have observations in one passband and mini-filaments may be too faint to be visible.\\
            \item[e] The spire indicates the top of the fan-like structure and the boundary indicates the lateral position of the fan-spine-like structure.
        \end{tablenotes}
    \end{threeparttable}
\end{table*}

\subsection{An example of jets driven by rising mini-filaments} \label{fila}

\begin{figure*}
\includegraphics[trim=0cm 3.2cm 0cm 0.5cm, width=0.9\textwidth]{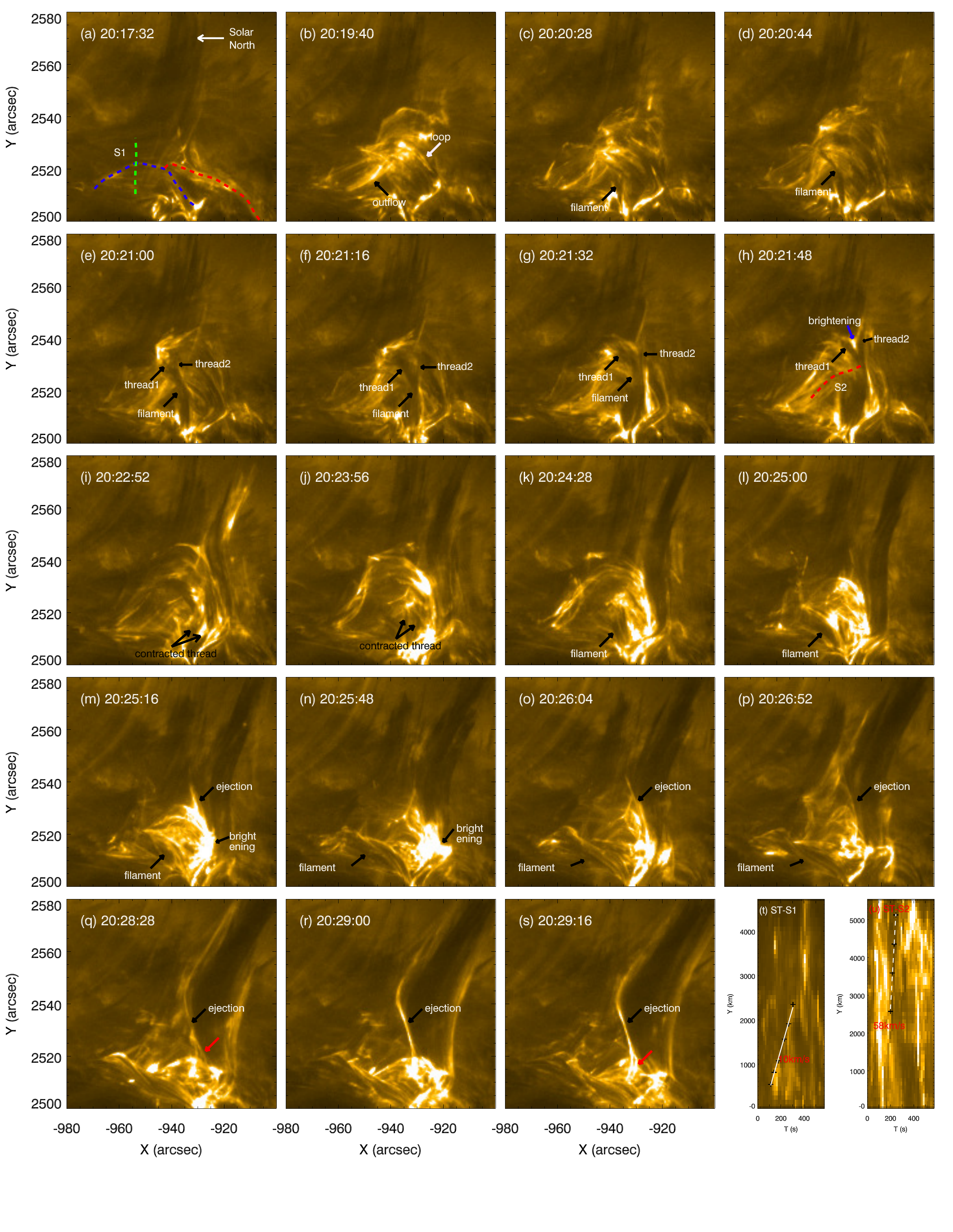}
\caption{Evolution of ``jet5''.
(a) - (s) The snapshots of EUI at 20:17:32\,UT, 20:19:40\,UT, 20:20:28\,UT, 20:20:44\,UT, 20:21:00\,UT, 20:21:16\,UT, 20:21:32\,UT, 20:21:48\,UT, 20:22:52\,UT, 20:23:56\,UT, 20:24:28\,UT 20:25:00\,UT, 20:25:16\,UT, 20:25:48\,UT, 20:26:04\,UT, 20:26:52\,UT, 20:28:28\,UT, 20:29:00\,UT and 20:29:16\,UT.
The blue and red dashed lines in (a) outline two sheared arcades.
The green dashed line ``S1'' in (a) shows the trajectory where we make the time-distance plot shown in (t).
The black and white arrows in (b) point to the outflow and post-reconnected loops.
The black arrows in (c) - (g) marked with ``filament'' points to the rising mini-filament.
The black arrows in (e) - (h) marked with ``thread1'' and ``thread2'' point to different threads of the rising mini-filament and the blue arrow in (h) points to the brightening at the top of the fan-like structure.
The red dashed line ``S2'' in (h) shows the trajectory where we obtain the time-distance plot shown in (u).
The black arrows in (i) - (j) point to the contracted threads.
The black arrows marked with ``filament'' in (k) - (p) point to the newly-formed mini-filament.
The black arrows marked with ``brightening'' in (m) and (n) point to the brightenings at the bottom of the jet.
The black arrows marked with ``ejection'' in (m) - (s) point to the ejections originating from the brightenings.
The red arrows in (q) and (s) point to the structures at the bottom of the ejections.
(t) and (u) show the time-distance plots along ``S1'' and ``S2''.
All the snapshots are rotated 90 degrees counter-clockwise as same as Figure\,\ref{fig:ove} (b).
\label{fig:ex1}}
\end{figure*}

The first example of jets is called ``jet5'' in Table \ref{tab:para}, which was facilitated at 20:21:48\,UT.
The activated area was confined by two sheared arcades initially (outlined by blue and red dashed lines in Figure \ref{fig:ex1} (a)).
The arcades rose at 20:16:44\,UT with speeds at around $10\,\rm km~s^{-1}$ (see the time-distance plot along ``S1'' in Figure \ref{fig:ex1} (t)).
The two rising arcades experienced a complex interaction process, which produced both bidirectional outflow and post-reconnected loops (pointed by black and white arrows in Figure \ref{fig:ex1} (b)).
The reconnection outcome was not directly related to the trigger of the studied jet, thus, its associated physical parameter and process will not be shown here.
Their confinement to the underlying  structure decreased during this process.
We observed that a mini-filament (marked by a black arrow in Figure \ref{fig:ex1} (c) - (g)) under these loops emerged at 20:20:28\,UT.
The mini-filament kept rising after it appeared and approached the spire of the fan-like structure.
We measured the rising speed of the mini-filament through the time-distance plot along ``S2'' (see the red dashed line in Figure \ref{fig:ex1} (h)) and found that it rose with a velocity of around $58\,\rm km~s^{-1}$ (see Figure \ref{fig:ex1} (u)).
We can trace the evolution of threads in the mini-filament with unprecedentedly-high-resolution observations (pointed by black arrows in Figure \ref{fig:ex1} (e) - (h)).
We observed that a brightening between these two threads appeared at the top of the fan-like structure (pointed by blue arrow in Figure \ref{fig:ex1} (h)), and plasma were ejected outward from there after that.
The southern thread (``thread2'' in Figure \ref{fig:ex1}) interacted with the fan-spine-like structure and produced ``jet5''.
The northern thread (``thread1'' in Figure \ref{fig:ex1}) contracted toward the bottom of the fan-spine-like structure.
We can observe the whole contracting process of these mini-filament threads, and the contracted threads were marked by black arrows in Figure \ref{fig:ex1} (i) \& (j).
The new mini-filament formed after the contraction of these threads, as pointed by black arrows marked with ``filament'' in Figure \ref{fig:ex1} (k) - (p).
We also notice that some brightenings appeared during this jet (pointed by black arrows marked with ``brightening'' in Figure \ref{fig:ex1} (m) \& (n)), and some bright plasma were ejected from there and formed bright threads in the jet (see black arrows marked with ``ejection'' in Figure \ref{fig:ex1} (m) - (s)).
The ejection also experienced complex interaction process as we noticed that the connectivity at the bottom changed during the ejections (see red arrows in Figure \ref{fig:ex1} (q) \& (s)).
This observation offers clear evidence of how the rising mini-filament rose, partially erupted and formed a new mini-filament, and these processes can occur repeatedly to drive recurrent jets.

\subsection{An example of jets driven by rising loops} \label{loop}

\begin{figure*}[ht!]
\includegraphics[trim=0cm 1cm 0cm 0.5cm, width=\textwidth]{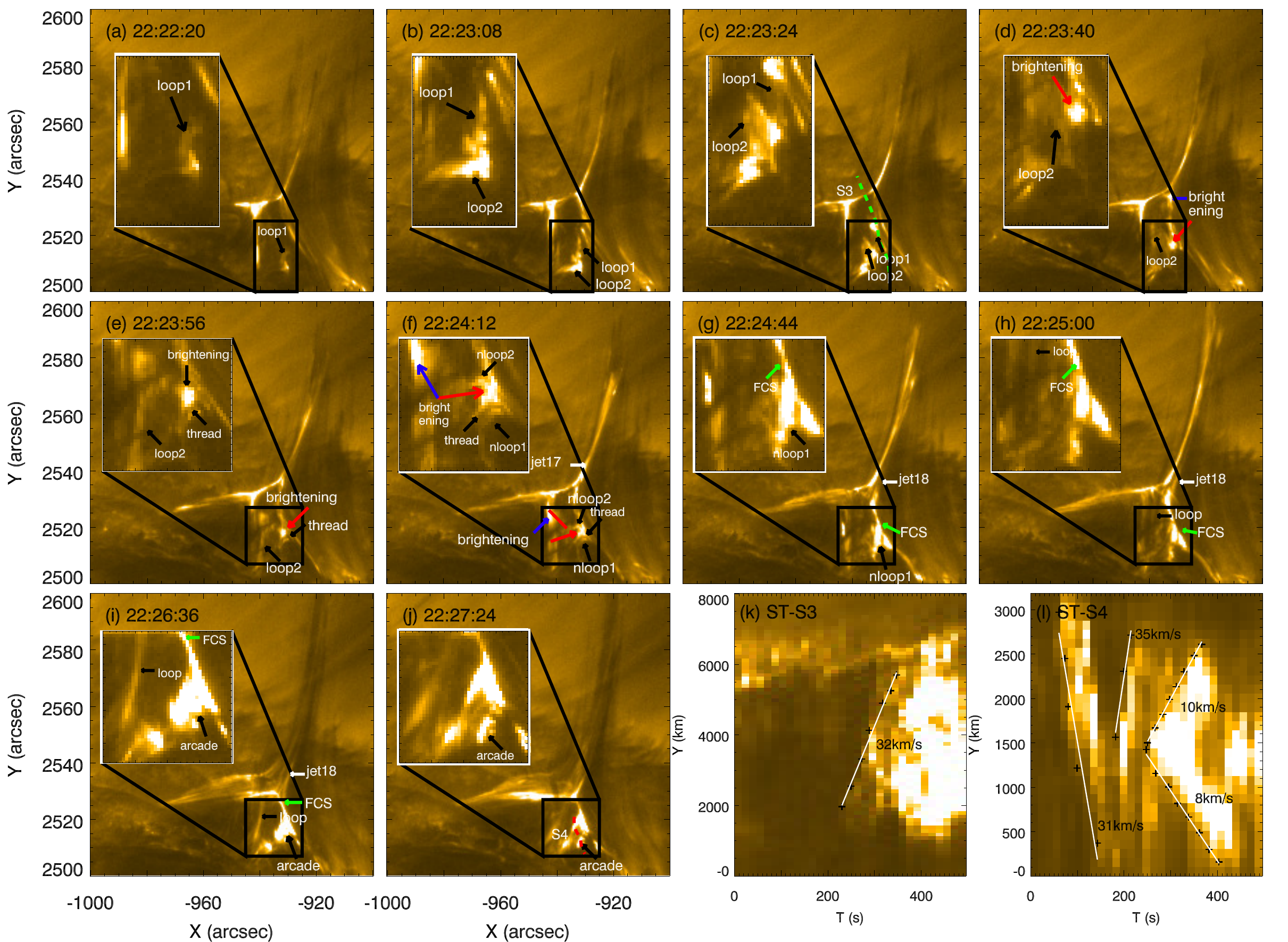}
\caption{Evolution of ``jet18''.
Panels (a) - (j) show the snapshots of EUI at 22:22:20\,UT, 22:23:08\,UT, 22:23:24\,UT, 22:23:40\,UT, 22:23:56\,UT, 22:24:12\,UT, 22:24:44\,UT, 22:25:00\,UT, 22:26:36\,UT and 22:27:24\,UT.
The black rectangles in all panels show the region where we enlarged.
The black arrow in (a) points to the first rising loop.
The black arrows in (b) - (e) point to the first and second rising loops.
The green dashed line ``S3'' in (c) shows the trajectory where we make the time-distance plot shown in (k).
The blue arrows in (d) \& (f) point to the brightenings occurred when the rising loops encountered the fan-spine-like structure.
The red arrows in (d) - (f) point to the brightening near the rising loop.
The black arrows marked with ``thread'' in (e) \& (f) point to the thread that appeared during the brightening.
The black arrows marked with ``nloop1'' and ``nloop2'' in (f) \& (g) point to loops that appeared during the brightening.
The red line in (f) shows the trajectory of the second rising loop.
The white arrow in (f) points to ``jet17''.
The black arrows in (h) \& (i) marked with ``loop'' point to the post-reconnected loops.
The green arrows in (g) - (i) point to the bright thin sheet which may be the flare current sheet in breakout model.
The white arrows in (g) - (i) point to ``jet18''.
The black arrows marked with ``arcade'' in (i) \& (j) point to contracted arcade under the FCS.
The red dashed line ``S4'' in (j) shows the trajectory where we make the time-distance plot shown in (l).
Panels (k) - (l) show the time-distance plots along ``S3'' and ``S4''.
All the snapshots are rotated 90 degrees counter-clockwise as same as Figure\,\ref{fig:ove} (b).
\label{fig:ex2}}
\end{figure*}

The second example of jets occurred at 22:24:44\,UT and is called ``jet18'' in Table \ref{tab:para}.
This jet was induced by the interaction between emerging loops and pre-existing fan-spine-like structure.
Before this jet occurred, a loop pointed by black arrows marked with ``loop1'' in Figure \ref{fig:ex2} (a) - (c) appeared at the bottom of the fan-spine-like structure, followed by the rising of the second loop (pointed by black arrows marked with ``loop2'' in Figure \ref{fig:ex2} (b) - (e)).
The first loop appeared at 22:22:20\,UT and rose with a speed of around $32\,\rm km~s^{-1}$, which was obtained by the time-distance plot (see Figure \ref{fig:ex2} (k)) along ``S3'' (marked by green dashed line in Figure \ref{fig:ex2} (c)).
The second loop appeared at 22:23:08\,UT, and its rising speed was estimated by its moving distance (outlined by red line in Figure \ref{fig:ex2} (f)) divided by the time it spent, which was at around $39\,\rm km~s^{-1}$.
These two loops approached the spire of fan-like structure, 
and we observed that brightenings appeared when the rising loops encountered the spire of the fan-like structure (pointed by blue arrows in Figure \ref{fig:ex2} (d) \& (f)).
The interaction between the rising loops and fan-spine-like structure produced several post-reconnected loops (pointed by black arrows in Figure \ref{fig:ex2} (h) \& (i)).
We also observed a brightening appeared near these rising loops (pointed by red arrows in Figure \ref{fig:ex2} (d) \& (f)), followed by the formation of a bright thin sheet (pointed by green arrows in Figure \ref{fig:ex2} (g) - (i)).
A small thread or mini-filament may participate in this process as we observed that a thread (pointed by black arrows marked by ``thread'' in Figure \ref{fig:ex2} (e) \& (f)) appeared before the brightening and disappeared after that.
This brightening process also produced two loops pointed by black arrows marked with ``nloop1'' and ``nloop2'' in Figure \ref{fig:ex2} (f) \& (g), and we can estimate the contracting speed of ``nloop1'' in the time-slice plot along``S4'', which is around  $35\,\rm km~s^{-1}$ (see Figure \ref{fig:ex2} (l)).
This ``jet18'' (pointed by white arrow in Figure \ref{fig:ex2} (g) - (i)) was ejected from one end of the bright thin sheet, and we observed contracting arcades (see black arrows marked with ``arcade'' in Figure \ref{fig:ex2} (i) \& (j)) at the other end of the bright thin sheet.
These arcades contracted with speeds at around $10\,\rm km~s^{-1}$, and the outflow region moved with a speed of around $8\,\rm km~s^{-1}$ toward the opposite direction (see Figure \ref{fig:ex2} (l)).
\citet{2026A&A...706A...1D} also performed a detailed analysis of this example for its fine dynamic process, and more details can be found there.
The interaction between loops (rising from the bottom or expelled from a loop system) and fan-spine-like structures may occur repeatedly during the recurrent jets.

\section{Discussion and Conclusions} \label{sec:rescon}

We performed a detailed observational study of a series of recurrent jets originating from the same fan-spine-like structure.
We totally compiled statistics of 22 jets in this event.
These jets had lengths at around $3 - 18.6\,\rm Mm$ with an average at $8.8\,\rm Mm$, and their widths were in the range of $0.3 - 3.6\,\rm Mm$ with an average at $0.86\,\rm Mm$.
These jets propagated with speeds of $25 - 186\,\rm km~s^{-1}$, and their average speed was at $80\,\rm km~s^{-1}$.
The physical parameters of our samples are consistent with those small-scale coronal jets reported in \citet{2023ApJ...943...24P} and \citet{2025A&A...702A.188N}.
We also roughly estimated the kinetic energy of these jets, which was at around $10^{21} - 10^{25}\,\rm erg$.

The dynamics in this kind of recurrent jets has been widely investigated in previous studies.
\citet{2015ApJ...815...71C} observed several recurrent and homologous jets at the boundary of an active region, and these jets were induced by continuous interaction between a satellite spot/moving magnetic features and pre-existing fields.
\citet{2023NatCo..14.2107C} studied the continuous activities in a null point configuration, and they found the evidence that the recurrent jets were triggered by persistent null point reconnection, which showed recurrent brightening at the null point.
\citet{2024ApJ...962L..38D} reported three homologous jets, which were trigged by mini-filament eruptions due to the continuous emergence of magnetic flux and injection of magnetic helicity into the fan-spine structure.
\citet{2024A&A...691A.198J} and \citet{2024ApJ...973...74K} provided fine chromospheric evolution under recurrent jets by observations from the Swedish 1-m Solar Telescope (SST) and 1.6-m Goode Solar Telescope (GST) with spatial resolution close to nearly 30 km per pixel and 50 km per pixel respectively, however, their coronal dynamics were still not studied in detail due to the limitation of resolution of AIA.
At present, the direct observations of fine dynamics during recurrent jets caused by interaction between a fan-spine structure and persistent rising structures are scarce.
In this event, we report how persistent rising loops/mini-filaments interact with the fan-spine-like structure and then trigger recurrent coronal jets with the unprecedentedly high-resolution observations, and show fine dynamics in these processes.

At the bottom of the fan-spine-like structure, we observed numerous small loops and mini-filaments (take Figure \ref{fig:ove} (b) for example).
These loops and mini-filaments were extremely active and we often saw the lifting and brightening of them.
In most samples of our studied jets, we observed that existed loops or mini-filaments were rising from the bottom of the fan-spine-like structure before the jets occurred, and their rising process and fine dynamics of interacting with overlying fan-spine-like structure can be easily found in the related animation of Figure \ref{fig:ove}.
In Table \ref{tab:para}, the rising speeds of these loops and mini-filaments ranged from $8\,\rm km~s^{-1}$ to $58\,\rm km~s^{-1}$, with an average at $27\,\rm km~s^{-1}$.
We found that many recurrent jets in this event were induced by interaction between the rising loops and the fan-spine-like structure.
The remaining jets were triggered by rising mini-filaments, while there existed several jets whose trigger process can not be identified through the data.
We found the detailed dynamics of these rising structures during their rise and interaction with overlying fan-spine-like structure, including the fine dynamics of mini-filament threads in partially erupting, contracting and forming a new mini-filament, the expelling process of threads from loop systems, the fine dynamics of loops when they interacted with the fan-spine-like structure and etc..
This result indicated that these recurrent jets were produced by the interaction between the rising loops/mini-filaments and fan-spine-like structure occurring repeatedly, thus, these persistent rising loops/mini-filaments play a vital role in producing homologous and recurrent jets.
In the mini-filament-related case, we found that the mini-filament first rose with a speed of $58\,\rm km~s^{-1}$, and it experienced partial eruption, with part of its threads being ejected as a part of the jet and the remaining part contracting and forming a new mini-filament.
In the loop-related case, we found that a brightening near the footpoint of rising loops formed after the rising loops interacted with overlying fan-spine-like structure, followed by the formation of a bright thin sheet, which may correspond to a current sheet.
Some post-reconnected loops formed when the brightening appeared, and one of these loops contracted at $35\,\rm km~s^{-1}$.
We observed obvious contracting arcades at the outflow region of the current sheet, which contracted at $10\,\rm km~s^{-1}$, while the whole outflow region moved at $8\,\rm km~s^{-1}$ toward the opposite direction.
We also found that 18 in 22 jets were ejected from the spire of the fan-like structure while only 4 in 22 jets occurred at the lateral boundary of the fan-spine-like structure.
This indicates that the reconnection prefers to happen at the spire of the fan-like structure.

During the interaction between the rising loops/mini-filaments and the fan-spine-like structure, we found that bright thin sheets formed in some cases.
Many bright blobs were produced in these bright sheets as observed in Figure \ref{fig:pla} (a) - (f).
These bright blobs may correspond to plasmoids produced in current sheets by tearing instability, like the phenomenon studied in previous literature \citep[e.g.][etc.]{2014PhPl...21j2102W,2016ApJ...827....4W,2017ApJ...841...27N,2019ApJ...885L..15K}.
These bright blobs propagated along the bright thin sheets with an average speed at $21\,\rm km~s^{-1}$, consistent with previous studies \citep[e.g.][etc.]{2013ApJ...777...16Y,2016ApJ...827....4W,2025A&A...702A.188N}, and their average width was at $296\,\rm km$.
The observing features of these blobs were consistent with the plasmoid-mediated reconnection studied in \citet{2022NatCo..13..640Y} and \citet{2025A&A...702A.188N}, and the indirect indicator of plasmoid-mediated reconnection in simulation by \citet{2025A&A...702A.188N} (the boomerang-like structure) was also clearly observed in our observations (see arrows in Figure \ref{fig:pla} (g) \& (h)).
The bright blobs in bright thin sheets showed different behaviour.
Some only propagated along the bright thin sheets while others ejected plasma parallel to the jets during their propagation (see Figure \ref{fig:pla} (d) \& (f) for example).
The former is consistent with the classic dynamics of plasmoids in jets while the latter shows some difference.
\citet{2014PhPl...21j2102W} showed that the plasmoids was produced through fragmenting the null region and forming localized flux ropes.
The twist of localized flux ropes will spread along the fields and be assimilated into the reconnection outflow \citep{2016ApJ...827....4W}.
In several samples observed here, plasma can be ejected from the bright blobs, which indicates there may exist some unresolved mechanisms for ejections of a coronal jet.

In the future, observations combined with both images and magnetic field with higher spatio-temporal resolution and from multiple perspectives can be helpful to fully understand the mechanism of persistent rising process of structures beneath a fan-spine structure and the associated fine dynamics of the recurrent magnetic reconnection that drive the recurrent coronal jets.

\section*{Acknowledgements}
\addcontentsline{toc}{section}{Acknowledgements}
We are grateful to the anonymous referee for his/her constructive comments and suggestions that improve the manuscript.
This work is supported by the National Key R\&D Program of China (2021YFA0718600), the National Natural Science Foundation of China (Nos. 42230203, 12333009, 42174201) and the Fundamental Research Funds for the Central Universities (KG202506).
D.Y.D. was supported by NSFC grant 12403065.
Solar Orbiter is a space mission of international collaboration between ESA and NASA, operated by ESA. The EUI instrument was built by CSL, IAS, MPS, MSSL/UCL, PMOD/WRC, ROB, LCF/IO with funding from the Belgian Federal Science Policy Office (BELSPO/PRODEX PEA 4000112292 and 4000134088); the Centre National d’Etudes Spatiales (CNES); the UK Space Agency (UKSA); the Bundesministerium für Wirtschaft und Energie (BMWi) through the Deutsches Zentrum für Luft- und Raumfahrt (DLR); and the Swiss Space Office (SSO).
We are grateful to the Solar Orbiter and EUI teams for their wonderful observations and published data.


\section*{Data Availability}

 The data used in this paper can be obtained publicly, and they can be found in \url{https://www.sidc.be/EUI/data/}.



\bibliographystyle{mnras}
\bibliography{reference} 





\bsp	
\label{lastpage}
\end{document}